\documentclass[12pt]{article}
\usepackage{color,amsmath,amssymb,exscale,epsfig}
\input epsf
\newcommand{\m}[1]{\marginpar{{\tiny *}} }

\def\bea{\begin{eqnarray}}
\def\eea{\end{eqnarray}}

\begin{document}
\oddsidemargin -0.3cm
\vspace*{-2.3cm}
\begin{flushright}
\normalsize{
SLAC-PUB-14964
  }
\end{flushright}

\hfill
\vspace{20pt}
\begin{center}
{\Large \bf  Improving top quark induced charge asymmetries at the LHC using $t\bar t$ transverse momentum} 
\end{center}

\vspace{15pt}
\begin{center}
\large{Ezequiel \'Alvarez$^{a,b,}$\footnote{sequi@unsl.edu.ar}}

\vspace{20pt}
\textit{$^a$ CONICET, INFAP and Departamento de F\'{\i}sica, FCFMN,\\ Universidad Nacional de San Luis \\
Av. Ej\'ercito de los Andes 950, 5700, San Luis, Argentina}

\textit{$^b$ SLAC National Accelerator Laboratory, 2575 Sand Hill Road, Menlo Park, CA 94025, USA}

\end{center}

\begin{abstract}
We study how the significance of top quark induced charge asymmetries at the LHC may be enhanced exploiting the $t\bar t$ total transverse momentum to enrich the fraction of quark fusion events in the sample.  We combine this variable with previous variables related to the boost and the invariant mass of the $t\bar t$ pair to find an optimum cut which maximizes the significance of the asymmetry when systematic and statistic errors are taken into account.  We find that including the $t\bar t$ transverse momentum in the analysis of the expected 2012 LHC data provides a considerable enhancement in the significance of the asymmetry.
\end{abstract}

\newpage

\setcounter{footnote}{0}
\section{Introduction}
With the beginning of the LHC era and what could be the discovery of the Higgs field \cite{higgs} predicted more than 40 years ago, the circle of the Standard Model (SM) is getting closed with all its particles below the $\sim 100$ GeV scale.  The end of this important period in high energy physics places us in the doors of the upcoming challenging stage: the search for New Physics (NP).  The long to come results that the LHC and many other current experiments will deliver us in the next years will unveil the mystery if NP was around the corner or if, on the other hand, the SM was a long held theory.

From a theoretical point of view, one expects to have NP at the TeV scale to understand the mechanism responsible for spontaneous symmetry breaking without need of fine tuning.  This is usually known as the hierarchy problem.  From the observational point of view, there are many hints of NP but with no precise clues on its energy scale.  Examples of these are the indirect evidence of dark matter, the large matter-antimatter asymmetry and the small neutrino masses among others.  

May be the most studied theory that could solve many of the theoretical and observational concerns of beyond the SM physics is supersymmetry.  However, the first LHC results seem not to favor it, at least in its low energy version \cite{susydown}.  There are, however, many other NP theoretical frames which would solve some of the existing theoretical and observational problems.  In general, theories which attempt to solve the hierarchy problem, modify the top quark sector to avoid the large quadratic divergences coming from the top loop of the Higgs two point function.  From this point of view, and in addition to its little exploration, top physics has been considered an important window for the search of NP for a long time \cite{top90s}.

In the last years Tevatron has produced and studied top quarks as never before.  The most striking result is the recent measurement of the forward-backward $t\bar t$ asymmetry by CDF \cite{fb}, which in the SM is predicted to be not null at next to leading order (NLO) \cite{gr1,smfb}, but the experimental result disagrees with the SM prediction at a $3.4 \sigma$ level for large invariant masses.  This result has motivated many theoretical works \cite{anomaly} which attempt to understand the anomaly while not entering in conflict with other observables.  With the last year shut down of Tevatron, the final answer for this top puzzle has relied on the LHC. 

The LHC, being the largest collider machine ever built, has a large capability of producing top quarks.  Insofar has already produced and studied a number of top quarks which is more than an order of magnitude than those at Tevatron.   However, being a symmetric collider ($pp$ instead of $p\bar p$) makes it harder to follow the path of Tevatron $t\bar t$ forward-backward asymmetry.  As a matter of fact, there is no way to build directly a forward-backward asymmetry at the LHC, since there is not a referent beam on which define an absolute forward or backward direction.  However, although there is no preferred forward direction at the LHC, it is possible to {\it induce} an asymmetry by defining a forward direction event by event.  This possibility has been studied since the 90's \cite{gr1,gr2} and it has been further explored in many different ways \cite{gr3,chinos,hewett,otros,mai,tao,as}.  Along this work, we call {\it induced asymmetries} to refer in general to all the different asymmetries defined in these works.  The main idea relying in the event by event definition of the forward direction is that if a valence quark of one of the protons is involved in the hard process of the event, then the $t\bar t$ pair has more probability of having its $z$-momentum in the same direction as the incoming quark.  The reason for this is that the parton distribution function (PDF) for the proton assures that a valence quark is more likely to have more $z$-momentum than any other parton.  This conclusion has been exploited to induce a $t\bar t$ asymmetry at the LHC using different kinematic variables which attempt to enhance the asymmetry. 

Since the main processes that contribute to the asymmetry in the SM@NLO\footnote{In the SM@NLO also the $qg\to t \bar t$ process contributes to the asymmetry but, since its cross section is less than $3\%$ of the total \cite{mcatnlo}, we neglect it in the reasoning. In any case, the reasoning concerning the quark content of the initial partons would serve partially also for this process.  We do not neglect this process in the numerical calculations.} and in most of the NP scenarios are those which consist of quark fusion, $q\bar q \to t\bar t$, in many of the works which induce an asymmetry they attempt to isolate quark fusion events over gluon fusion events to enhance the asymmetry.  In most of the cases \cite{gr3,chinos,hewett,otros,mai,tao,as} this enhancement is again sought by aiming the fact that a valence quark is likely to carry more momentum than other partons.  In some of these works they also use polarization, spin correlation and relative rapidity of the tops \cite{mai,tao} and the $t\bar t$ transverse momentum, $p_T(t\bar t)$ \cite{gr3}.

In this article we exploit a recent variable which has only\footnote{After this work was completed, CMS published a note \cite{cmsnew} where they use $p_T(t\bar t)$ in their analysis.} been explored in Ref.~\cite{gr3}.  This consists in attempt to recognize gluon fusion events by using that these have higher probability to emit initial state radiation than quark fusion events.  Therefore, the $t\bar t$ pairs produced through gluon fusion are more likely to have higher transverse momentum.  Along this article we analyze how the transverse momentum of the $t\bar t$ pair may be used to enhance the induced $t\bar t$ asymmetries by discarding events which are likely to be gluon fusion.  We explore its possible combination with other pre-existing variables to optimize the perspectives for measuring the SM@NLO asymmetry at the LHC and, eventually, also a NP asymmetry.  In complement to the studies in Ref.~\cite{gr3}, our analysis is performed on a mathematical frame which relates the significance of the asymmetry to the capability of this new variable to increase the quark fusion content of the sample. We consider the interplay between enhancing the absolute value of the asymmetry and the systematic or statistic dominance of the error.  Our final comparison of the performance of using the $p_T(t\bar t)$ variable is done against the total significance of the asymmetry.  

The control of a new variable which helps to recognize gluon fusion events is useful to enhance the absolute value of the induced asymmetries. In fact, since at the LHC the induced asymmetries result largely diluted by the gluon fusion events, it could be possible that its absolute value lies below the systematic error, in which case its measurement would be precluded even with infinite events to study.  By introducing a simple model for systematic errors, we analyze this possibility in this article.

This article is divided as follows.  In the following section we study some general features concerning the significance of induced asymmetries at the LHC.  In section \ref{sect3} we use the $t\bar t$ transverse momentum variable to distinguish gluon from quark fusion events and apply it to the SM and to a NP benchmark model to see how the significance is improved.  In section \ref{remarks} we give our conclusions and final remarks.


\section{Significance of an induced asymmetry at the LHC}
\label{facts}

In this section we summarize some relevant results concerning the induction of an asymmetry in a symmetric collider as the LHC.  For an analysis in depth of diagnosis and comparison of general and particular proposals to induce an asymmetry in a $pp$ collider the reader should refer to \cite{jay}.

At the LHC, in contrast to the Tevatron, it is not possible to construct directly a forward-backward asymmetry in $t\bar t$ production events, since there is no preferred direction in the laboratory system.  However, as it has been studied in the last years \cite{gr1,gr2,gr3,chinos,hewett,otros,mai,tao,as}, it is possible to induce an asymmetry by setting an algorithm which indicates when an event should add to $N_+$ and when to $N_-$ (see below).  Usually, this algorithm is a function of the top and anti-top kinematic variables in each event.  In general, the algorithm adds to $N_+$ when attempts to recognize that in a quark fusion event, $q\bar q\to t \bar  t$, the $z$-momentum of the outcoming top in the center of mass system is in the same direction as the incoming quark; and adds to $N_-$ when this $z$-momentum is in the same direction of the incoming anti-quark.

The induced asymmetry is defined as
\bea
A=\frac{N_+ - N_-}{N_+ + N_-},
\eea
where the precise definition of $N_\pm$ depends on which algorithm is being used to induce the asymmetry.  We will restrict in our analysis to $N_\pm$ as excluding variables, since it is shown in \cite{jay} that any case can be reduced to the excluding case and that the significance of the asymmetry is always enhanced when this is done.  

The measuring of the asymmetry carries an error which consists of systematic ($\sigma_{syst}$) and statistic ($\sigma_{stat}$) parts.  Along this work we consider these errors independent, therefore the total error ($\sigma$) is the quadrature sum of both.  We define the significance of the asymmetry as its deviation from zero in units of its error,
\bea
{\cal S} = \frac{A}{\sigma} .
\eea
One may also define the systematic and statistic significance by using their corresponding error; which will prove to be useful in the different regimes where one of the errors dominates over the other.  

The reason for defining the significance of the asymmetry as its deviation from zero is the following.  We consider in this work two cases of asymmetries: {\it i)} the SM@NLO asymmetry, where we are interested in measuring the NLO effects which produce the deviation from zero of the asymmetry; and {\it ii)} the NP asymmetry, which would produce a deviation from the SM@NLO expected value.  In the second case, however, we work at tree level (or leading order, LO) since it is possible to show \cite{darold} that if both asymmetries are small then they may be added linearly.  Henceforth, also in the second case the NP effect would be a deviation from the SM@LO expected value, which is zero.

It is interesting to study how the asymmetry and its significance are modified as a cut is performed on the sample and the number of total events and the fraction of quark fusion events change.  In fact, suppose that we have a first sample of $N$ events of $t\bar t$ production from which a fraction $f_{q1}$ comes from quark fusion and a fraction $1-f_{q1}$ from gluon fusion (see footnote 1).   Since in average the gluon fusion events do not contribute to the numerator of the asymmetry we have that the gluon fusion events reduce the value of the asymmetry by increasing the denominator.  In fact, if we define $\epsilon_1$ as the ideal asymmetry that would be measured if we could isolate only the quark fusion events in the sample, then is easy to see that the relationship between this thought asymmetry and the measured one is  \cite{jay} \footnote{Notice that we are not taking into account the fraction of events within $f_{q1}$ in which the algorithm fails to recognize which proton put the quark and which the anti-quark, since at this level this fraction may be taken as approximately constant and, hence, will not modify our conclusions.  To also include effects on how this fraction affects the result see \cite{jay}.}:
\bea
A_1 &=&\epsilon_1 \,f_{q1}.
\eea
Id est, the ideal asymmetry built from the quark fusion events is diluted because of $f_{q1}<1$.

If we now perform a cut with a given criteria on this first sample in such a way that the sample ends up with $f_s\, N$ selected events, from which a new fraction $f_{q2}$ comes from quark fusion, then the asymmetry with this second sample will be
\bea
A_2 &=& \epsilon_2 \, f_{q2}.
\eea
Where we have set $\epsilon_2 \neq \epsilon_1$, since the selection criteria usually produces a preferred selection in the quark fusion events, yielding a change in the quark fusion asymmetry.  This change, which we call $Q=\epsilon_2/\epsilon_1$, is usually an enhancement ($Q\geq 1$) and is model dependent.

In sight of next section analysis, it is interesting to study in this language how the significance of the asymmetry behaves in the limit of statistic and systematic error.  The statistic regime will be when the number of events is such that the statistic error dominates over the systematic one, whereas the systematic regime will be when the number of events is such that the statistic error is negligible.  In general, for the luminosity accumulated at the LHC insofar, we should be in the systematic regime.  However, given the strong cuts proposed in next section, the features of a statistic regime are important as well.

\begin{figure}
\begin{center}
\includegraphics[width=.55\textwidth]{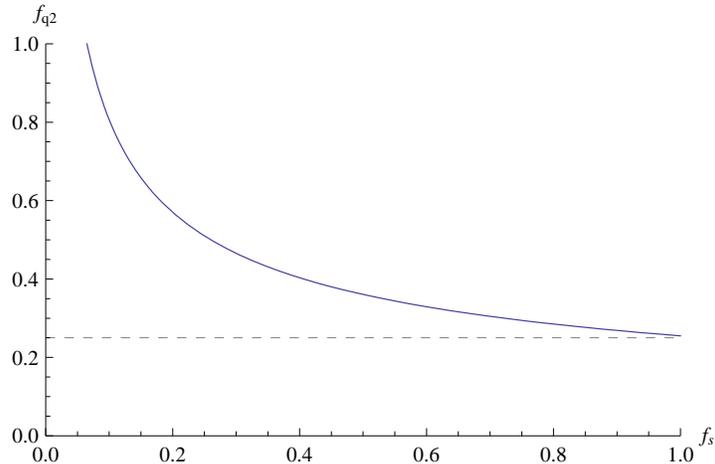}
\end{center}
\caption{Minimum requirement for the fraction of quark-fusion events ($f_{q2}$) in the selected sample in order to have an statistical successful cut for $Q=1$.  $f_s$ corresponds to the fraction of the original sample that is selected by the cut.  The horizontal dashed line indicates the original fraction of quark fusion events in the LHC@7TeV.}
\label{contour}
\end{figure}

We first study the statistic regime.  From the previous results we may see that, considering $N_\pm$ as binomial random variables, the ratio between the statistical significances of the first and second sample is given by \cite{jay}
\bea
{\cal S}_{stat\,2}/{\cal S}_{stat\,1} &=& Q \,\frac{f_{q2}}{f_{q1}} \sqrt{f_s} .
\eea
By setting this relationship equal to 1, we find the minimum value of $f_{q2}$ required to have a successful cut from the statistical point of view.  This value will depend not only on the selected fraction $f_s$, but also on the initial fraction $f_{q1}$ and the ratio $Q$ in which the quark fusion asymmetry is enhanced by the selection.  We plot in Fig.~\ref{contour}, for $Q=1$ and as a function of the selected fraction $f_s$, the minimal value of $f_{q2}$ required to have a successful statistical cut.  We take the original sample to be from LHC@7TeV, in which the fraction of quark fusion events in $t\bar t$ production is $f_{q1}\approx0.25$.  The minimum requirement for $f_{q2}$ shown in Fig.~\ref{contour} is very hard to achieve in the practical examples, and this is why most of the selection cuts do not improve the statistical significance (see discussion on Fig.~\ref{fractions} below), unless there is an enhancement coming from the $Q$ factor.

The $Q$ factor should be treated separately in each case, since its range of variation depends on the features of the cut and the model.  If the cut does not make a kinematic selection which points to enhance the quark fusion asymmetry, then we may find that $Q$ is usually between $1$ and $2$.   If, on the other hand, the cut selects the quark fusion events which enhance the asymmetry of a given model, then we may expect higher values of $Q$.  An example of the latter would be selecting high invariant mass events in a resonant NP model.  

In a systematic regime, on the other hand, the amount of data accumulated is such that the previous conclusions are no longer valid.  Since the systematic error depends on the precise observable, the region of kinematic space of each event and the measuring apparatus among others, it seems quite difficult to draw general conclusions in this case.  However, for the purposes of a simplified analysis, one may model the systematic errors as a fixed proportion of the variables that are constructed with the measurements.  In this case, we may work as in \cite{as} and model a systematic error to the number of events measured for each term of the asymmetry as $\sigma_{syst}(N_\pm) = c\,N_{\pm}$, with $c\approx0.01$ which is in concordance with the experimental results \cite{atlas1,cms1,cmsnew,atlasnew}.  From here one finds that in this model the systematic error of the asymmetry in this regime is constant and of order $c$.

It is important to notice at this point that if the absolute value of the asymmetry to be measured is smaller than $\sim 5 c$ then it will be impossible to distinguish it from zero.  We therefore see that in this regime any cut that increases the absolute value of the asymmetry it is successful, as far as the number of events does not decrease to the level of making important the statistic error.  Id est, any cut that increases the fraction of quark fusion events in the selected sample is valuable.  Therefore, the analogous to Fig.~\ref{contour} in a systematic error regime, would be a horizontal line at $f_{q2}=f_{q1}$.  

From the previous paragraphs we conclude that if in a given sample within a systematic error regime the asymmetry is not large enough, then it is crucial to induce a larger asymmetry by making intelligent cuts. The limit on these cuts comes from not decreasing the number of events below the systematic error regime, unless the cuts are specially designed to produce an important enhancement in the asymmetry (the $Q$ factor).

In next section we analyze some existing cuts and algorithms to induce an asymmetry in $pp$ collisions, and study an alternative kinematic variable which proves to be useful in discarding gluon fusion events and, therefore, enhancing the asymmetry.

\section{Analysis of selecting events using $t\bar t$ transverse momentum}

\label{sect3}
In this section we analyze the efficiency to enhance an induced asymmetry and its significance at the LHC by using new algorithms based on the total transverse momentum of the $t\bar t$ system.  

The inquiry to induce asymmetries at the LHC has begun since the conception of a symmetric machine \cite{gr1,gr2} and has taken special force with the first measurements in the last years \cite{gr3,chinos,hewett,otros,mai,tao,as}.   Most of the proposed induced asymmetries consist in a cut to enhance the quark fusion fraction in the sample and an algorithm to attempt to recognize which proton put the quark, if is it the case that was quark fusion.  Although these are all different proposals, there is a general reasoning underlying in most of them for the cut and for the algorithm.  The main idea in the cut is that in the proton a valence quark is likely to have considerably more momentum that any other anti-quark, hence the quark fusion events are more likely to be boosted.  And the general underlying idea in the algorithm to recognize which proton put the quark is that since the quark has in general more momentum that the anti-quark (which is always from the sea), then the $z$-direction of the $t\bar t$ pair in the lab system is likely to be the same as the incoming quark. Based on these main ingredients, the different proposals look for an enhancement of the asymmetry and its significance using different methods.

For instance, the first proposed {\it Central charge asymmetry} \cite{gr1,gr2} would look for a difference in the number of tops and anti-tops that would come out in a central region of rapidity limited by a cutting rapidity $y_C$.  This proposal was advantageous in the sense of having many events, but disadvantageous in having too many gluon fusion events which dilute the asymmetry.  Hence, it is more useful in a statistic regime rather than in a systematic one \cite{jay}.  A second kind of example is the one that looks for boosted $t\bar t$ pairs \cite{chinos,as}, which has the convenience of reaching a high quark fusion fraction, but with the fault of making strong cuts which may rise the statistic error above the desired limit.  These proposals use different variables as a measure of the the $t\bar t$ boost; for instance and for future reference, in \cite{as} they use
\bea
\beta&=&\frac{|p^z_t + p^z_{\bar t}|}{E_t +E_{\bar t}}
\label{beta}
\eea
to indicate how boosted is the $t\bar t$ pair.  $\beta=0$ indicates no boost and $\beta\to1$ indicates maximum boost.  This variable instead of $|p^z_t + p^z_{\bar t}|$, besides of being experimentally more robust \cite{as}, turns out to be less correlated to $m_{t\bar t}$ and hence makes less relevant the dependence on $Q$.

The above referred induced asymmetries have explored in depth the many possibilities that come out from the underlying main idea of the large momentum carried by a valence quark.  Moreover, some of the works have also proposed and explored new ideas to enhance the asymmetry, as for instance \cite{mai,tao} using top polarization, spin correlation and relative rapidity and \cite{gr3} using the transverse momentum of the $t\bar t$ pair, which is the variable we study in the following paragraphs.

\subsection{Improving the quark fusion fraction through $t\bar t$ transverse momentum}

We explore in this section an alternative idea to recognize quark fusion events which relies on the total transverse momentum of the $t\bar t$ pair, $p_T(t\bar t)$.  This alternative tool will allow us to improve the final result of the quark fusion content of the sample and, therefore, to improve the significance of the asymmetry in the systematic error regime.  The main idea underlying here is the higher probability of initial-state radiation between initial gluon hard scattering compared to initial quark hard scattering.  This can be seen directly from the QCD Lagrangian color factors appearing in gluon and quark scattering and its ratio is $9/4$.
~ Although the details of these processes lie beyond the scope of this work, the interested reader may refer to studies in Higgs production with high transverse momentum through gluon fusion \cite{deflo}, which have the same radical basis.  Hence, although the underlying idea has many years, it has only been used in Ref.~\cite{gr3} in $t\bar t$ production, as far as we are concerned (see footnote 2).

\begin{figure}
\begin{center}
\includegraphics[width=.8\textwidth]{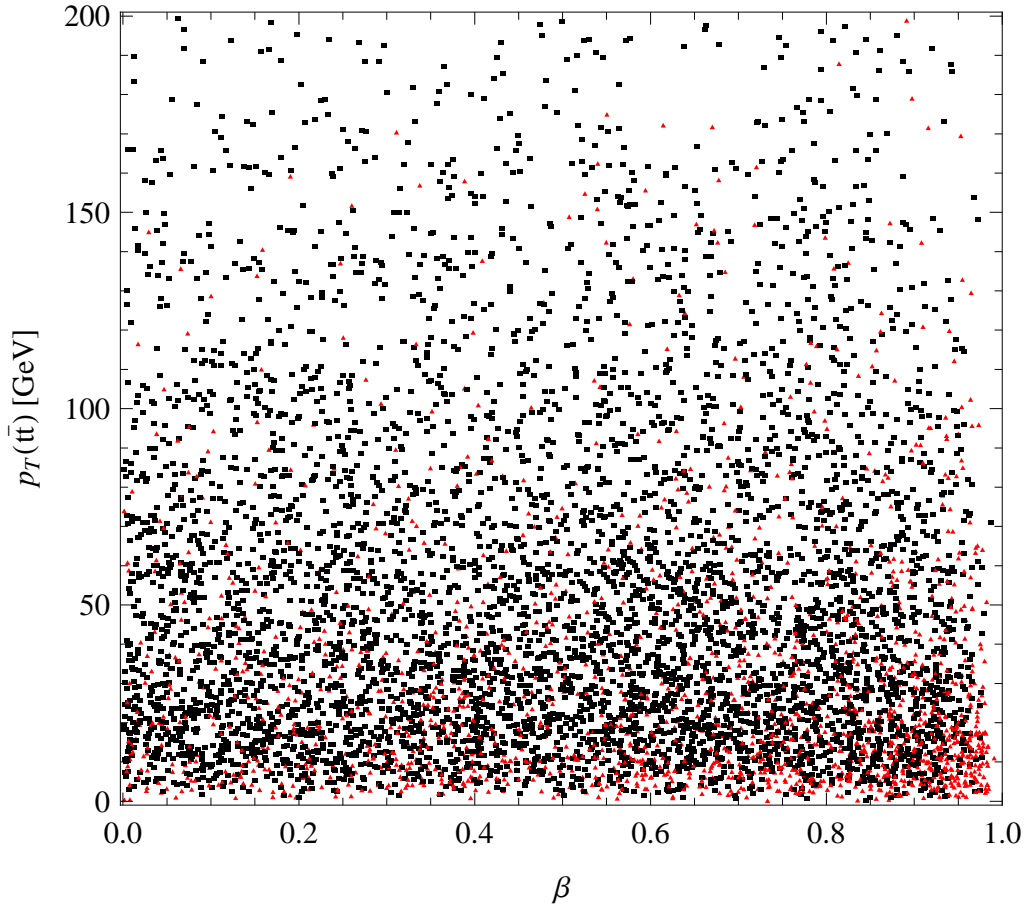}
\end{center}
\caption{[color online] Initial state partons in $t\bar t$ production as a function of the total transverse momentum and boost of the $t\bar t$ pair (see Eq.~\ref{beta}).  Black points (squares) correspond to gluon fusion events and red points (triangles) to quark fusion events.  The percentages of gluon fusion events in the regions $p_T<20$ GeV, $20\ \mbox{GeV}\leq p_T <50$ GeV, $50\ \mbox{GeV}\leq p_T <100$ GeV and $100$ GeV$\leq p_T$ are 56\%, 76\%, 85\% and 91\%, respectively.}
\label{transverse}
\end{figure}

As an illustrative exploration of the sought effect we perform a numerical analysis of the initial state radiation using Pythia \cite{pythia}.  These results could be improved by more precise calculations.  In order to see numerically the effect of the initial state radiation on the transverse momentum of the $t\bar t$ pairs as a function of the initial partons, we have simulated $t\bar t$ production through MadGraph/MadEvent \cite{mgme} and passed them through Pythia. The simulation was done for LHC@7TeV using the Standard Model.  In Fig.~\ref{transverse} we plot $10\,000$ events as a function of $p_T(t\bar t)$ and $\beta$ (see Eq.~\ref{beta}), using different colors according to the initial state partons of the event.  As it can be seen from the figure, there is the expected correlation between the higher transverse momentum events and the gluon fusion initial state (black points).  Moreover, one can also see in the figure the well known correlation between the high $\beta$ events and the quark fusion events (red points).

With the purpose of increasing the size of the asymmetry in a systematic error regime, we are interested in isolating regions of kinematic variables in which a large fraction of quark fusion events is present.  Moreover, since as cutting the sample the statistic error increases --which could spoil the systematic error regime--, we are also interested in exploring how the quark fusion content of the sample is modified as the sample is gradually cut.  With this goal we have essayed different cutting methods while counting the quark fusion fraction of the sample.  

\begin{figure}[!htb]
\begin{center}
\includegraphics[width=.65\textwidth]{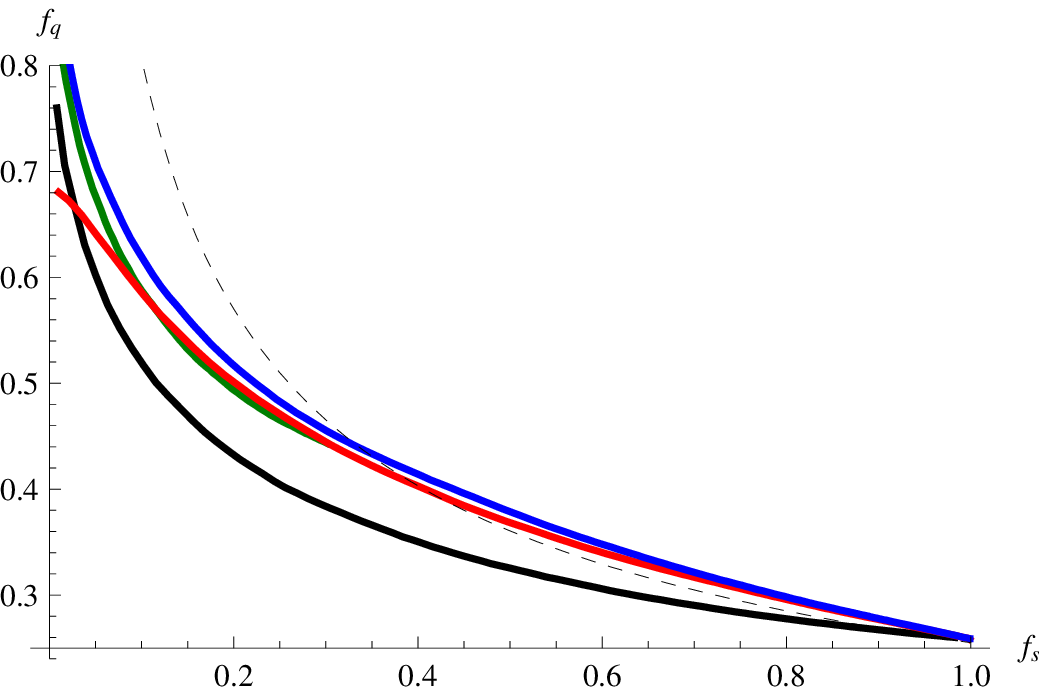}
\end{center}
\caption{[color online] Quark fusion fraction as a function of selected fraction for different methods for gradually selecting the sample.   Black solid line represents a cut using only $\beta$, while the other solid lines, which improve the quark fusion content, also use $p_T(t\bar t)$ in their cut.  See text for the detailed explanation of each solid line. The solid lines at $f_s=0.2$ from bottom to top are black, green, red and blue, respectively. The dashed line corresponds to the statistic limit in Fig.~\ref{contour} and its reference value is only relevant in a statistic regime analysis, as describe in the text. } 
\label{fractions}
\end{figure}

Based on Fig.~\ref{transverse}, we first cut the sample by sweeping a vertical line from $\beta=0$ to $\beta=1$ (selecting the events with $\beta$ larger than the line), in order to gradually isolate a large quark fusion fraction sample.  For instance, when the vertical line is in $\beta=0.2$ we select the events with $\beta>0.2$ and count which is the fraction of events that have been selected ($f_s$) and which is the quark fusion fraction ($f_q$) in the selected sample.  This kind of selection corresponds to the original cut in \cite{as} and the result is plotted in solid black in Fig.~\ref{fractions}. Notice that, in contrast to Ref.~\cite{as}, we plot $f_q$ against the selected fraction, since we consider it the relevant variable for this article. 

As a second essay we sweep a horizontal line from top to bottom in Fig.~\ref{transverse}, which would mean to only take into account the transverse momentum of the $t\bar t$ pair, neglecting a correlation with $\beta$.  The result is plotted in solid red in Fig.~\ref{fractions} and, as it may be seen, this new method reaches better quark fraction content than the previous one.  

As a third possibility we try a cut that combines both variables.  Given that most of the quark fusion events are in the right lower corner of the plot in Fig.~\ref{transverse} we find suitable to sweep a straight line with slope $50$ GeV from top to bottom.  The resulting quark fusion fraction is plotted in solid blue in Fig.~\ref{fractions}.  We see that the quark fusion fraction as a function of the selected fraction is improved comparing it to the previous essays.  As a matter of fact, this is approximately the best cut we could perform.  We have tried different shapes and swipes and we have not found a considerable improvement to this cutting method.


At last, we sweep Fig.~\ref{transverse} from left to right with a vertical line, but with the additional constraint of $p_T(t\bar t)<20$ GeV.  This is plotted in green in Fig.~\ref{fractions}.  Of course that in this case, due to the overall cut in $p_T(t\bar t)$, the selected fraction reaches a maximum of $f_s \approx 0.3$.  This kind of cutting method will prove to be useful below to compare the significances as a function of $\beta$ for different values of cut in $p_T(t\bar t)$.

The results in Fig.~\ref{fractions} give a clear picture on how the different cutting methods behave in selecting quark fusion events and how the new variable in game, $p_T(t\bar t)$, gives a variety of possibilities to enhance induced asymmetries.  First, we should notice that in a statistic regime is very hard to place a cut that increases the significance, id est to have its fraction $f_q$ above the dashed line for some value of $f_s$, unless there is an enhancement coming from $Q$.  Observe that this conclusion is independent of the total number of events as far as being in a statistic regime.  Second, in a systematic regime, the solid lines in the figure tell us that all the cuts will enhance the asymmetry.  However, this will occur as far as the total number of events in the selected sample is large enough to do not spoil the systematic regime by increasing the statistic error.  In any case,  as previously mentioned, the final comparison of the different cuts needs the additional ingredient of the $Q$ factor for each model and cut.  Henceforth, it is suitable at this point to compare some different cutting methods from the significance point of view.

\subsection{Significance comparison using $t\bar t$ transverse momentum}

In order to verify the improvement in the significance of the asymmetry that can be reached using $p_T(t\bar t)$, we have selected a concrete induced asymmetry and probed our improvements in the SM@NLO and in a benchmark model of NP.  We have chosen as an algorithm to attempt to recognize which proton put the quark, the same as Refs.~\cite{chinos,as}.  This means that we add 1 to $N_+$ when $|y_t|-|y_{\bar t}|>0$, and vice-versa.  For the NP model, we have chosen a positive point from \cite{pheno} which we have tested to pass at a 95\% C.L.~the CDF results \cite{fb} and the new constraints from ATLAS \cite{atlas1,atlasdijets,atlasnew} and CMS \cite{cms1,cmsnew}. To be more precise, this point refers to a 1 TeV axigluon with coupling to right tops $f_{t_R}=7 g_s$, to light quarks $f_{q_R}=-0.13 g_s$, and all other couplings equal to zero. ($g_s$ refers to the QCD coupling constant.)  The resulting width that comes out from these couplings is $\Gamma\approx 400$ GeV, and therefore our simulations use running width. Given the large coefficient of the top coupling, we have computed the coefficients of the partial wave amplitudes that contribute to the $q\bar q \to t\bar t$ process using this point of parameter space and found no violations of unitarity constraints.  Notice also that most of the solutions to the forward-backward asymmetry with heavier intermediate particles need larger light quark couplings and, therefore, have been rejected by the dijet NP searches \cite{atlasdijets}.

\begin{figure}[!htb]
\begin{center}
\hspace*{-1.4cm}
\begin{minipage}[b]{0.42\linewidth}
\begin{center}
\includegraphics[width=1\textwidth]{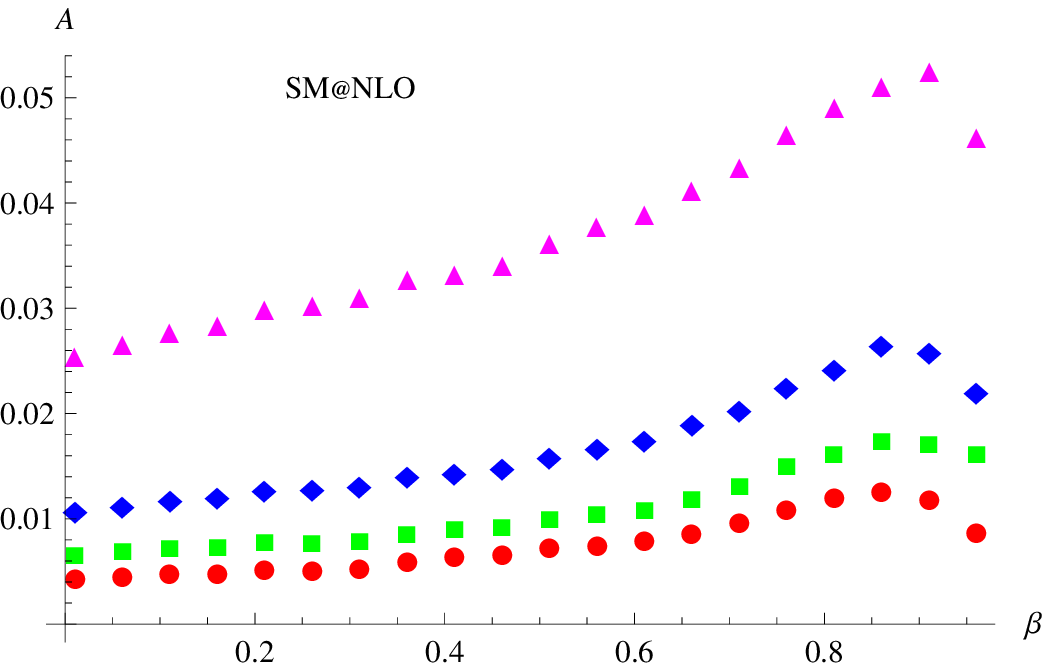}
\newline
(a)
\end{center}
\end{minipage}
\hspace{.3cm}
\begin{minipage}[b]{0.42\linewidth}
\begin{center}
\includegraphics[width=1\textwidth]{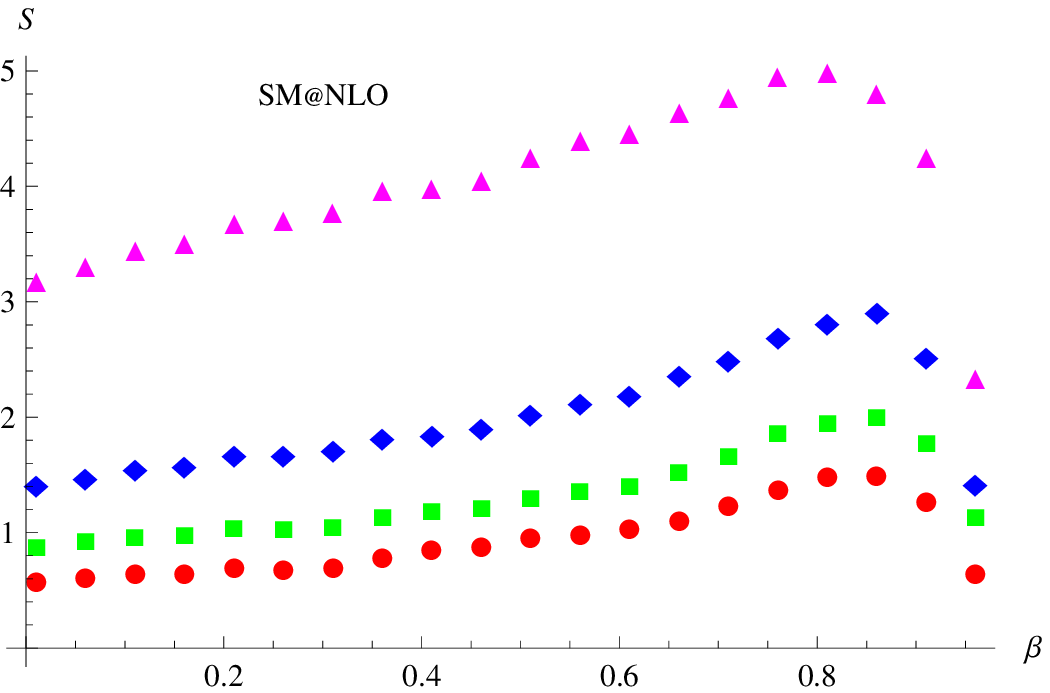}
\newline
(b)
\end{center}
\end{minipage}
\end{center}
\begin{center}
\hspace*{-1.4cm}
\begin{minipage}[b]{0.42\linewidth}
\begin{center}
\includegraphics[width=1\textwidth]{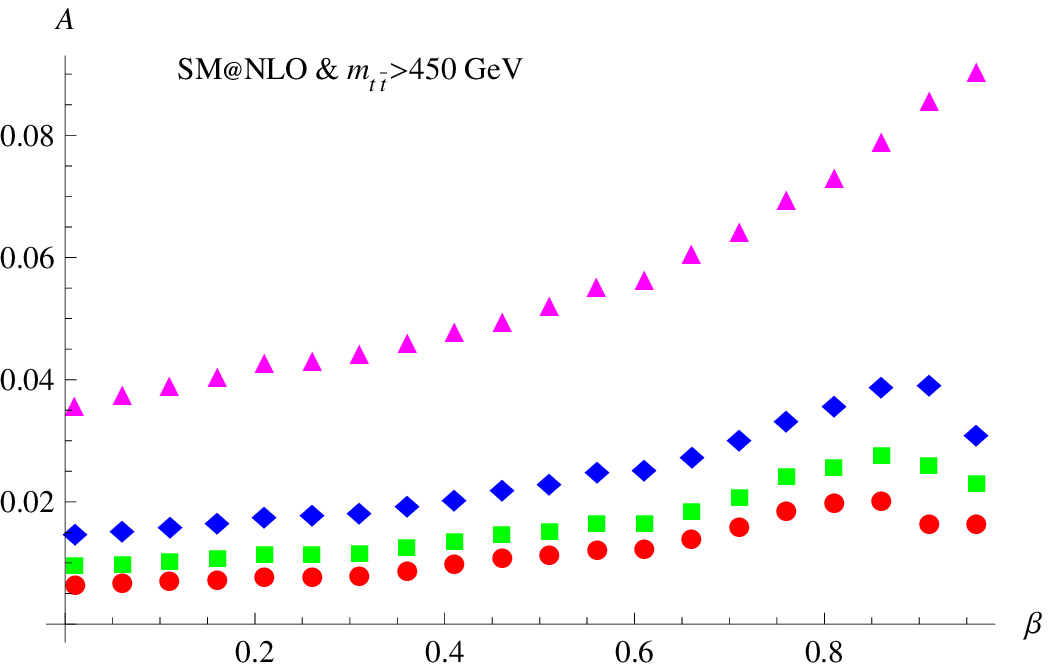}
\newline
(c)
\end{center}
\end{minipage}
\hspace{.3cm}
\begin{minipage}[b]{0.42\linewidth}
\begin{center}
\includegraphics[width=1\textwidth]{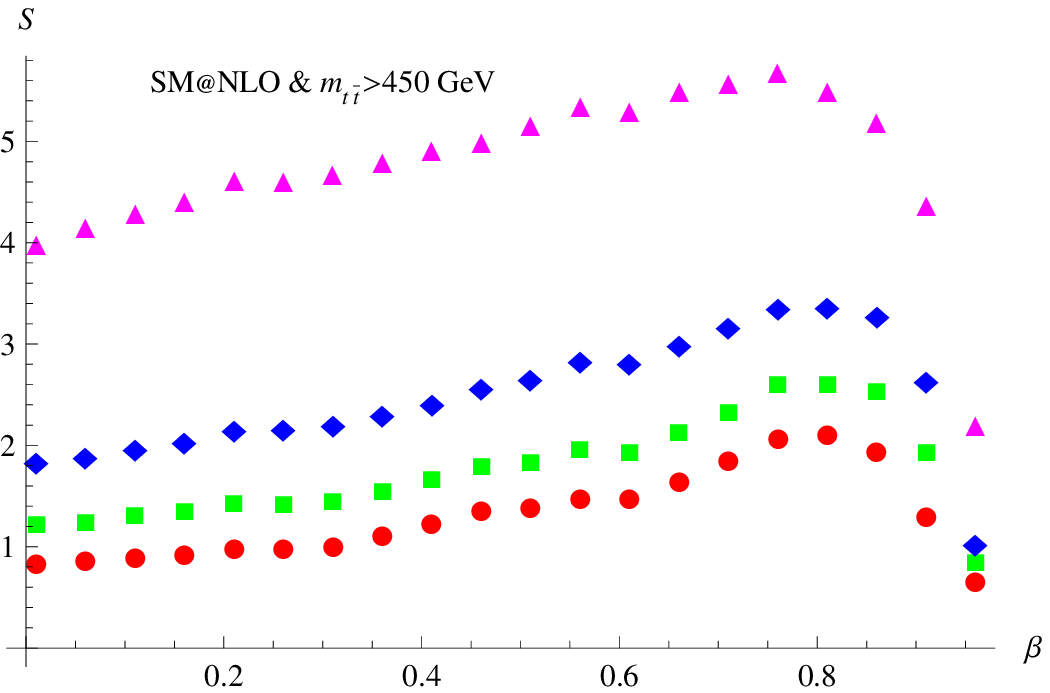}
\newline
(d)
\end{center}
\end{minipage}
\end{center}
\caption{[color online] Asymmetry and its significance as a function of the boost ($\beta$) for different cuts in $p_T(t\bar t)$ for the SM@NLO sample expected by the end of 2012 assuming a $6\%$ selection efficiency.  The systematic error is parameterized as a $c=0.01$ fraction of $N_\pm$.   Bottom panels include an additional cut in the invariant mass, $m_{t\bar t}>450$ GeV.  From bottom to top,  the different plots correspond to no cut in $p_T(t\bar t)$ (red circles), $p_T(t\bar t) <  100 \mbox{ GeV (green squares), } 50 \mbox{ GeV (blue diamonds) and }20 \mbox{ GeV (magenta triangles)}$. Id est, red points correspond to not using the $t\bar t$ transverse momentum, whereas the other plots show the improvement in using such a kinematic variable.} 
\label{significance}
\end{figure}

We have used MC@NLO \cite{mcatnlo} showered with Herwig \cite{herwig} to simulate de SM@NLO events and MadGraph/MadEvent with Pythia to simulate the SM+NP tree level events.  In both cases we have simulated $5 \mbox{fb}^{-1}$ of LHC@7TeV and $15 \mbox{fb}^{-1}$ of LHC@8TeV, which is the expected \cite{lum} accumulated luminosity by the end of 2012.   From these simulated events we have assumed an average selection efficiency of $6\%$, which would correspond to the analyzed channels in Ref.~\cite{cmsnew}.  As a simplified model \cite{as} for the systematic errors we have used a constant fraction of the counting $N_\pm$ variables, $c=0.01$, which is consistent with the systematic errors in Ref.~\cite{cmsnew}.  We have not decayed the top quarks, therefore our simulation lacks of a final state analysis through reconstruction of tops and detector simulation.   

\begin{figure}[!htb]
\begin{center}
\hspace*{-1.4cm}
\begin{minipage}[b]{0.42\linewidth}
\begin{center}
\includegraphics[width=1\textwidth]{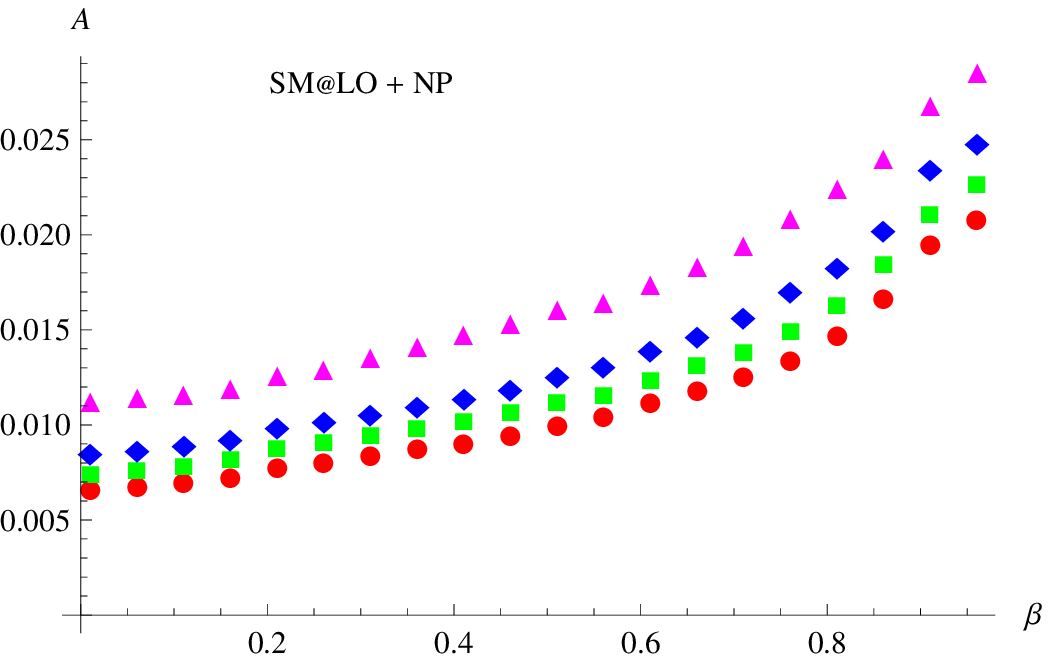}
\newline
(a)
\end{center}
\end{minipage}
\hspace{.3cm}
\begin{minipage}[b]{0.42\linewidth}
\begin{center}
\includegraphics[width=1\textwidth]{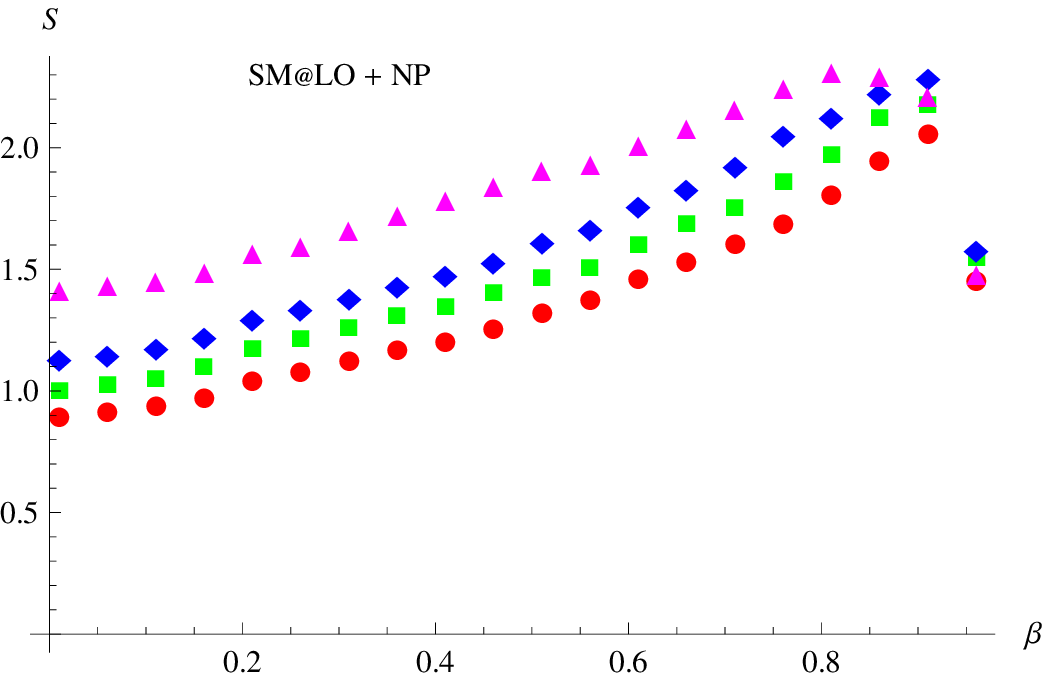}
\newline
(b)
\end{center}
\end{minipage}
\end{center}
\begin{center}
\hspace*{-1.4cm}
\begin{minipage}[b]{0.42\linewidth}
\begin{center}
\includegraphics[width=1\textwidth]{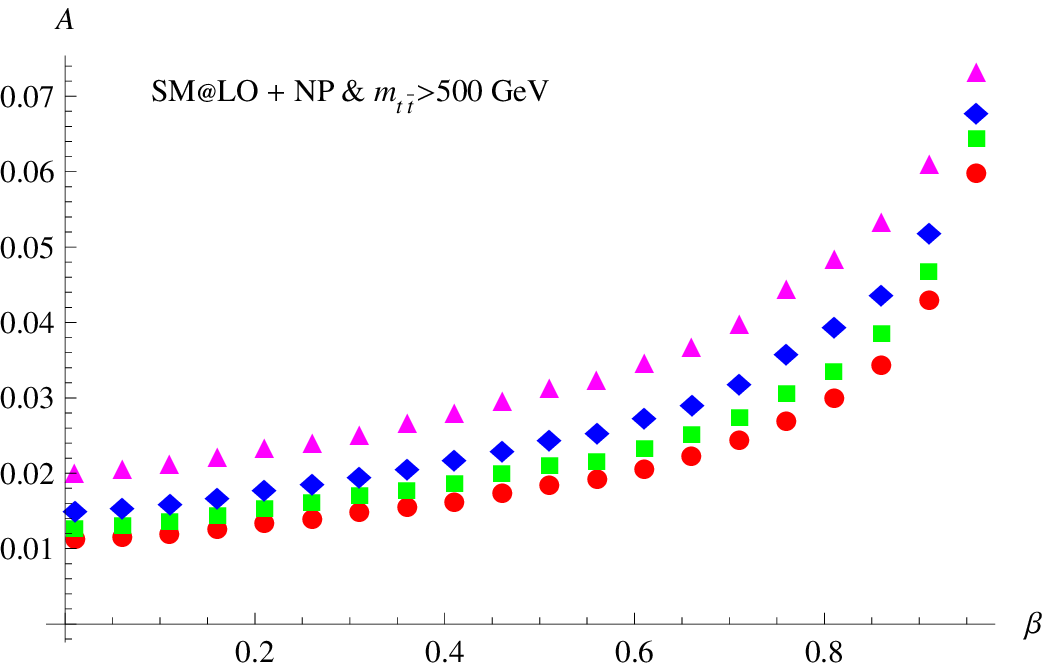}
\newline
(c)
\end{center}
\end{minipage}
\hspace{.3cm}
\begin{minipage}[b]{0.42\linewidth}
\begin{center}
\includegraphics[width=1\textwidth]{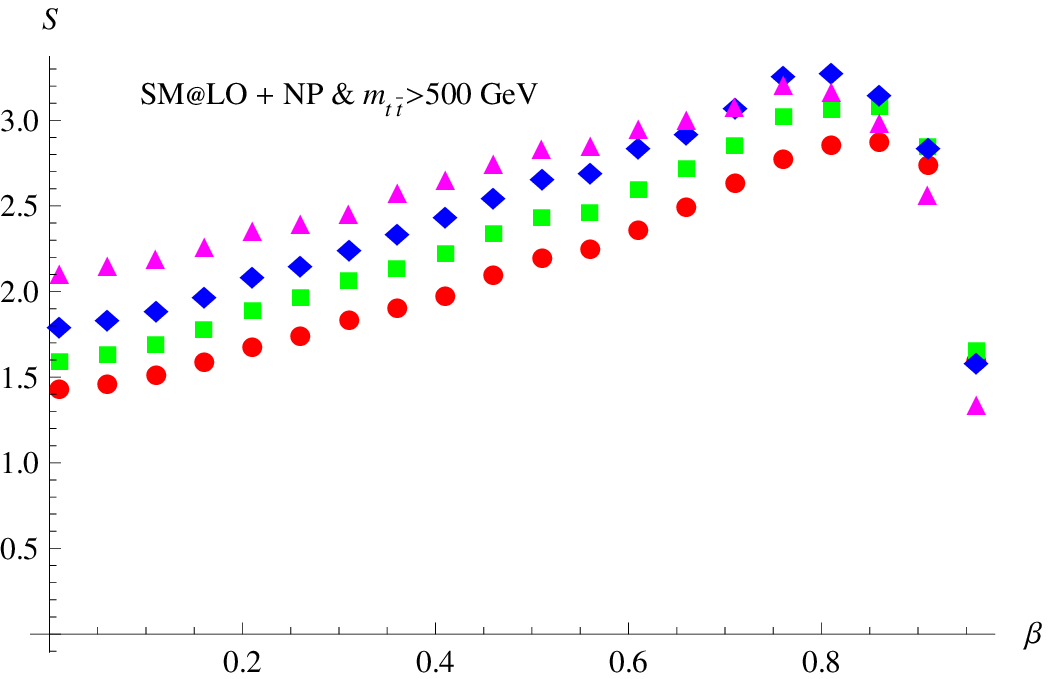}
\newline
(d)
\end{center}
\end{minipage}
\end{center}
\caption{[color online] Idem as Fig.~\ref{significance}, but for the SM@LO+NP sample using the benchmark model of a $1$ TeV axigluon.  The NP model is quite conservative and produces a NP forward backward asymmetry at the Tevatron of $A_{FB}=0.04$.}
\label{significance2}
\end{figure}

In order to make a consistent comparison with previous works \cite{as}, we have gradually selected the sample by varying the parameter $\beta$, and placed a different cut in $p_T(t\bar t)$ for each case.  We have plotted the results in Fig.~\ref{significance} for the SM@NLO case, and in Fig.~\ref{significance2} for the SM@LO+NP case. The left panels of both plots correspond to the asymmetry and the right panels to its significance.  In the top rows there are no additional cuts, whereas in the bottom rows an additional cut in the invariant mass is performed.  

The cut in the invariant mass produces an enhancement in the fraction of quark fusion events, but also a diminishing in the statistic of the sample.  Both of these depend on the center of mass energy available in the $pp$ collision.  To obtain simple conclusions, we have used the same invariant mass cut in both center of mass energy samples (LHC@7TeV and LHC@8TeV). We have found that for these specific samples a cut in $m_{t\bar t}>450$ GeV and $m_{t\bar t}>500$ for the SM@NLO and SM@LO+NP examples, respectively, optimizes the maximum in the significance.  However, a discriminant cut for different center of mass energies and luminosities for each example is recommended for a detailed analysis. 

As it can be seen from the figures, some plots have a decrease in the slope for large values of $\beta$.  In the left plots (asymmetries) this is due to a relative decrease of the asymmetry for high values of $\beta$, as was already noticed in Ref.~\cite{as}.   In the right plots (significances) there is an additional reason for a decrease in the significance that comes from the growth of the error.  In fact, since at that level the cut is so strong that only a few thousand events are left in the sample, there is a considerable growth of the statistic error as $\beta \gtrsim 0.8$.  Notice in Figs.~\ref{significance2}c and \ref{significance2}d how the points with a $20$ GeV cut (magenta) in $p_T(t\bar t)$ fall below those with a $50$ GeV cut (blue) for $\beta\gtrsim 0.8$ because of this growth.

We also point out that we have used a quite conservative benchmark NP model which produces little asymmetry at the LHC ($A=0.007$ with no cuts at 8TeV) and at Tevatron ($A_{FB}=0.04$) due to its small couplings to light quarks and large width. However, we have verified that the enhancement in the significance is also valid in the same or better proportions with other not so conservative models.

From the plots in Fig.~\ref{significance} and Fig.~\ref{significance2} we see that the incorporation of the transverse momentum of the $t\bar t$ pair in the analysis provides an enhancement in the significance of the asymmetry.  In addition, we have verified that if more statistic would be accumulated, then the enhancement would increase still more for the cases where the invariant mass cut is applied.  For instance, the point with maximum significance in Fig.~\ref{significance}d would raise from $5.6$ to $10.3$ if only systematic errors would be retained.  

Finally, it is worth to point out that we have explored different combinations in the three variables $\beta,\ m_{t\bar t}$ and $p_T(t\bar t)$ in order to seek for an optimum enhancement in the significance.  The plots with maximum significance presented here correspond to the best solutions we have found.  Notice also that including $p_T(t\bar t)$ in the analysis helps to avoid the experimental issues that appear in measuring events in the $\beta \gtrsim 0.9$ region, while reaching still higher significances.  Moreover, we have verified that a combined analysis of the three variables improves the results of analyzing separately each one of them.  In fact, for instance for the SM@NLO case, if only the $p_T(t\bar t)$ would be used then the maximum significance would be ${\cal S}\approx3.2$ (Fig.~\ref{significance}b), if only $\beta$ would be used then the maximum significance would be ${\cal S}\approx 1.5$ (Fig.~\ref{significance}b), and if only $m_{t\bar t}$ would be used then the maximum significance would be ${\cal S}\approx 1.1$ (not shown in figures).  These significances should be compared to the maximum significance reached in Fig.~\ref{significance}d which is ${\cal S}\approx 5.6$.

\section{Final remarks}

\label{remarks}
In this article we have shown how the $t\bar t$ transverse momentum may be used to recognize and discard gluon fusion events and, therefore, enhance at the LHC the $t\bar t$ induced asymmetries and their significances.  

We have studied how to implement cuts on a raw $t\bar t$ sample in order to increase the quark fusion fraction of the selected sample.  We have shown which is the limit relationship between the relevant variables in which a given cut is convenient from the statistical point of view.  On the other hand, we have shown that any cut that increases the quark fusion fraction of the sample is convenient if we are in a systematic error regime.  As a matter of fact, the only limitation to a cut in a systematic regime comes from not increasing the statistic error to the level of the systematic error.

We have verified that the usage of the $t\bar t$ transverse momentum improves previous selection methods which aim to increase the quark fusion fraction content of the selected sample.

Using the LHC sample expected by the end of 2012 for the SM@NLO and the SM@LO+NP and assuming a selection efficiency of $6\%$, we have shown that using cuts in the $t\bar t$ transverse momentum may provide a gain in the significance of the induced asymmetry which may vary from $0.5$ to $3$ points in the best of the cases.  These absolute increasings correspond to large relative enhancements of the original significance.

Our analysis has been performed at parton level and, therefore, lacks of detector simulation and top reconstruction.  We expect that a complete analysis will lower the value of the significance of the asymmetry, but there is a priori no reason to lower the relative enhancement produced by the inclusion of $p_T(t\bar t)$ in the analysis.  

In a full analysis it would be recommendable to use different cuts in invariant mass for the different center of mass energy samples.  Also a cut which isolates the bottom right triangle (instead of rectangle) in Fig.~\ref{transverse} would improve the results.  However, this depends on the $t\bar t$ transverse momentum resolution that could be achieved.  It is worth noticing that, independent on the transverse momentum resolution and the exact theoretical prediction for its distribution, the cut should be as strong as possible until it reaches the level where the statistic error begins to lower the significance.  We would also like to stress that, although this work shows how to enhance the significance of one observable by discarding events, when chasing a specific model in a full analysis one should use, with different weights, all the events in different observables.  In fact, although there are some events which are more likely to come from gluon fusion than others, every event has information on the sought interaction which should be exploited.

Finally, from the experimental point of view, we conclude that a combined differential analysis of the three variables $\beta$, $m_{t\bar t}$ and $p_T(t\bar t)$ would be more fruitful that a separated differential analysis in each of them \cite{cmsnew}.  Also, given the very high resolution reached in $p_T(t\bar t)$ \cite{cmsnew}, the accumulation of luminosity will have a considerable positive impact on the significance of the induced asymmetries.

\section*{Acknowledgments}
Thanks to M.~Peskin for pointing out the initial state radiation idea and S.~Hoeche for setting up the MC@NLO resource.  Thanks to D. De Flori\'an for useful conversations and J.~Hewett and A.~Szynkman for reading the manuscript.  Thanks Conicet for the special funding and SLAC for its charming hospitality.
{}

\end{document}